\documentstyle[prd,aps,epsf,eqsecnum]{revtex}
\tighten
\draft 

\begin{document}

\title{Approximating the inspiral of test bodies into Kerr black holes}

\author{Kostas Glampedakis$^{1}$, Scott A.\ Hughes$^{2}$, and
Daniel Kennefick$^{3}$}
\address{$^{1}$Department of Physics and Astronomy, Cardiff
University, P.\ O.\ Box 913, Cardiff CF24 3YB, UK}
\address{$^{2}$Institute for Theoretical Physics, University of
California, Santa Barbara, CA 93103, USA}
\address{$^{3}$Theoretical Astrophysics, California Institute of
Technology, Pasadena, CA 91125, USA}

\date{\today}

\maketitle

\begin{abstract}
We present a new approximate method for constructing gravitational
radiation driven inspirals of test-bodies orbiting Kerr black holes.
Such orbits can be fully described by a semi-latus rectum $p$, an
eccentricity $e$, and an inclination angle $\iota$; or, by an energy
$E$, an angular momentum component $L_z$, and a third constant $Q$.
Our scheme uses expressions that are exact (within an adiabatic
approximation) for the rates of change ($\dot{p}$, $\dot{e}$,
$\dot{\iota}$) as linear combinations of the fluxes ($\dot{E}$,
$\dot{L_z}$, $\dot{Q}$), but uses quadrupole-order formulae for these
fluxes.  This scheme thus encodes the exact orbital dynamics,
augmenting it with approximate radiation reaction.  Comparing inspiral
trajectories, we find that this approximation agrees well with
numerical results for the special cases of eccentric equatorial and
circular inclined orbits, far more accurate than corresponding
weak-field formulae for ($\dot{p}$, $\dot{e}$, $\dot{\iota}$).  We use
this technique to study the inspiral of a test-body in inclined,
eccentric Kerr orbits.  Our results should be useful tools for
constructing approximate waveforms that can be used to study data
analysis problems for the future LISA gravitational-wave observatory,
in lieu of waveforms from more rigorous techniques that are currently
under development.
\end{abstract}



\section{Background and motivation}
\label{sec:background}

The capture of stellar-mass compact objects by massive black holes
residing in galactic nuclei is expected to be one of the most
important sources of gravitational radiation for the future LISA
space-based detector {\cite{lisa,rates}}.  Observing such events will
provide information about stellar dynamics in galactic nuclei, and
should make possible precise measurements of black holes' masses and
spins.  Indeed, the waves generated by such a capture will encode a
detailed description of the black hole's spacetime, making it possible
to test whether the ``large object'' in the galactic nucleus is indeed
a Kerr black hole as predicted by general relativity, or is some
exotic massive compact object {\cite{ryan_multipoles,hughes_proc}}.

Extracting such information will require accurate modeling of the
gravitational waveform.  The smallness of the system's mass ratio
(typically, $\mu/M \sim 10^{-4} - 10^{-6}$, where $\mu$ and $M$ are
the masses for the captured body and the central hole respectively)
allows one to treat the small body as a ``test particle'' moving in
the gravitational field of the black hole.  In the absence of
radiation, the small body moves on a geodesic orbit of the black hole
{\cite{chandra}}.  These orbits have three integrals of motion (apart
from $\mu$): energy $E$; angular momentum projected on the hole's spin
axis, $L_z$; and Carter's third constant $Q$, related to the square of
the angular momentum projected onto the equatorial plane.  A body in a
generic (eccentric and inclined) Kerr orbit traces an open ellipse
precessing about the black hole's spin axis, resulting in a
complicated overall motion.  Astrophysical captured bodies will move
in such complicated orbits.

The integrals of the motion are not constant in the presence of
gravitational radiation --- they evolve as energy and angular momentum
are carried away by the waves.  Due to the small mass ratio, they
should change adiabatically, on timescales much longer than any
orbital timescale.  Hence, the orbit looks geodesic on short
timescales.  This fact can be used to calculate gravitational
perturbations induced by the particle at infinity and at the hole's
event horizon, using the Teukolsky-Sasaki-Nakamura formalism
{\cite{teuk}}.  In this way, one can explicitly find the gravitational
waveform at infinity and compute the corresponding fluxes of $E$ and
$L_z$ to infinity and into the hole.  If one also knows the evolution
of the Carter constant, then the adiabatic nature of the inspiral
allows one to treat the small body's motion as an evolution through a
sequence of orbits: the body's worldline ${\bf z}(t)$ is that of a
geodesic orbit ${\bf z}_{\rm geod}(t)$ whose orbital constants are
slowly changing:
\begin{equation}
{\bf z}(t) = {\bf z}_{\rm geod}[t; E(t), L_z(t), Q(t)]\;.
\label{eq:adibatic_inspiral}
\end{equation}
Computing the inspiral properties is reduced to computing the
parameter space trajectory $[E(t)$, $L_z(t)$, $Q(t)]$.

One can in fact infer the change in $Q$ and thus fix the small body's
inspiral in two special cases: orbits that are equatorial, and orbits
that are circular but inclined.  A considerable amount of effort has
been devoted to studying these orbits and their evolution due to
gravitational-wave emission {\cite{rr,cutler,chapter,scott1,scott2,kgdk}}.
In these special cases, the evolution of the Carter constant $Q$ is
constrained: it remains constant at $Q = 0$ (equatorial orbits) or
evolves such that the system's eccentricity is constant at $e = 0$
(circular, inclined orbits) {\cite{ryan1,ori,minothesis}}.  Accurate
numerical computations, based on extracting $\dot E$ and $\dot L_z$
from fluxes of gravitational waves to infinity and down the event
horizon, have detailed the effects of radiation reaction and the
nature of gravitational-wave emission in these cases.

Unfortunately, this ``flux-balancing'' prescription fails in general
--- there is no known method for computing the rate $\dot{Q}$ from the
gravitational-wave fluxes in the absence of special constraints.  At
the moment, the only applicable result is a weak-field,
quadrupole-order calculation by Ryan {\cite{ryan1}}, who used a
weak-field radiation reaction force to infer $\dot{Q}$.  Not
surprisingly, Ryan's results become increasingly inaccurate and
unreliable as the orbit comes closer to the black hole.  It is likely
that a strong-field gravitational self-force prescription will be
needed to compute $\dot{Q}$.  Many groups are now working on this
problem {\cite{force}}.  It is generally acknowledged that no result
applicable to strong-field Kerr orbits should be expected within the
next few years.  In the meantime, therefore, an investigation of
possible approximation schemes for describing radiation reaction and
wave emission by these orbits is highly desirable.  Such schemes will
play an important role in mapping out the scope of the data analysis
task that the LISA community faces, making possible a realistic
assessment of issues such as the amount of computing power needed, the
accuracy with which black hole characteristics can be measured, and
the difficulty of measuring signals if the inspiral rate is large
enough to create a confusion-limited background {\cite{sterl}}.

In the remainder of this paper, we present such an approximate scheme.
The essential idea is to use the exact Kerr black hole geodesics to
describe the system's dynamics, but to evolve through a sequence of
those geodesics using the weak-field quadrupole-order fluxes for $\dot
E$ and $\dot L_z$.  Because this scheme mixes an exact notion of
short-timescale motion with an approximate description of the
long-timescale radiation effects, we call it a ``hybrid''
approximation.  We find that the hybrid approximation faithfully
reproduces features seen in the numerical strong-field analyses of
radiation reaction.  For example, we find that the rate of change of
eccentricity will typically switch sign prior to plunging into the
black hole; as a consequence, the orbit has substantial eccentricity
near the end of inspiral.  Self-consistent leading order calculations
(which approximate the orbital dynamics as well as the radiation
reaction) strongly underestimate this residual eccentricity.  In some
cases, they predict that the orbit is circular at the end of inspiral.
This incorrect circularization could have a big effect on the waveform
models that are used to lay the foundations of LISA data analysis,
since a circular inspiral produces waves with less interesting
harmonic structure than eccentric inspirals.  We advocate this hybrid
scheme as a method that is simple enough to produce waveforms that are
``fast and dirty'', but accurate enough to qualitatively reproduce
features that should exist in real inspirals.  We emphasize that the
hybrid waveforms are {\it not} the ultimate models one would want to
use as templates for analyzing the LISA datastream.  Instead, we
advocate them as tools for exploring issues in LISA data analysis, as
described in the paragraph above.

The ideas behind the hybrid approximation and key equations are given
in Sec.\ {\ref{sec:method}}; some of the more cumbersome details are
presented in Appendix {\ref{app:formulae}}.  In Sec.\
{\ref{sec:compare}, we then compare this technique's predictions to
those of detailed numerical calculations for the two special cases
that are well-understood now, equatorial orbits and inclined, circular
orbits.  We compare with the leading order results developed by Ryan
{\cite{ryan1}}, and show that the hybrid scheme qualitatively recovers
features seen in the strong-field numerical calculations.  Our results
show that holding the inclination angle $\iota$ constant is more
accurate than letting it evolve in the way that the weak-field fluxes
``want'' it to evolve (as compared to strong-field numerical
calculations).  We argue in Sec.\ {\ref{sec:compare}} and Appendix
{\ref{app:almostsphere}} that this tells us that the ``gravitational
potential'' felt by the inspiraling body is nearly spherical, and
argue further that holding $\iota$ constant should work well for
arbitrary orbits.

In Sec.\ {\ref{sec:generic}}, we move to ``generic'' configurations,
studying inspirals through a sequence of inclined, eccentric orbits.
In most cases, we find that the inspiral trajectories are
qualitatively similar to inspiral in the equatorial plane.  In
particular, we find that most configurations plunge into the black
hole with substantial residual eccentricity.  We also map out the
range of parameter space where we do not trust the hybrid
approximation: when orbits reach too deeply into the strong field, or
spiral in near inclination $90^\circ$, the weak-field fluxes that we
use do not appear to be reliable.  We conjecture in Sec.\
{\ref{sec:zoomwhirl}} on how an approximation could be developed to
better understand the Carter constant's evolution.  This approximation
is based on the ``zoom-whirl'' behavior of strong-field eccentric
orbits, recently described in Ref.\ {\cite{kgdk}}.  We provide
concluding discussion and suggest directions for future work on this
problem in Sec.\ {\ref{sec:conclude}}.  Throughout this paper, we use
units in which $G = c = 1$.


\section{The hybrid approximation}
\label{sec:method}

Generic Kerr geodesics can be parameterized by a triplet of constant
orbital elements: the semi-latus rectum $p$, the eccentricity $e$, and
the inclination angle $\iota$.  The elements $p$ and $e$ define the
orbit's radial turning points, the apastron and periastron:
\begin{equation}
r_a= \frac{p}{1-e}, ~~~~ r_p= \frac{p}{1+e}\;.
\label{turn_points}
\end{equation} 
In the strong field of a Kerr black hole, there are many ways that one
could define an ``inclination angle'' --- for example, the turning
points of the orbit's latitudinal motion, or the angle at which the
small body crosses the equator as seen by distant observers.  We use
the following definition:
\begin{equation}
\cos\iota= \frac{L_z}{\sqrt{Q + L_z^2}}\;.
\label{ioteq}
\end{equation}
This definition does not correspond to either of these examples, but
is very convenient: it depends simply on orbital constants and has a
useful intuitive description, suggesting that the Carter constant $Q$
is essentially just the square of the angular momentum projected into
the equatorial plane. (This description is in fact exactly correct for
Schwarzschild black holes; for non-zero spin it is not quite correct,
but is good enough to be useful.  We discuss this issue in more detail
in Appendix {\ref{app:almostsphere}}.)  The orbital elements can be
written as functions of $(E, L_z, Q)$, and vice versa.  Consequently,
we can write their time-derivatives as ${\dot p} = {\dot p}(p, e,
\iota, {\dot E}, {\dot L_z}, {\dot Q})$, and similarly for ${\dot e}$
and ${\dot\iota}$.

As already mentioned, we do not yet know how to accurately calculate
$\dot{Q}$.  It is only known to leading order in $M/p$ and in the spin
of the black hole {\cite{ryan1}}.  The orbital parameters used in
reference {\cite{ryan1}} are an eccentricity $\bar{e}$ (different from
$e$), a semi-major axis $\bar{a}$, and an inclination angle $\iota$
(identical to our $\iota$).  The two sets of parameters are related by
\begin{eqnarray}
1-e^2 &=& (1-\bar{e}^2) \left [ 1- \frac{4a}{M} \left 
(\frac{M}{p} \right)^{3/2} e^2 \cos\iota  \right]\;,
\\
p &=& \bar{a} (1-e^2) \left [ 1- \frac{2a}{M} \left 
( \frac{M}{p} \right )^{3/2}
e^2 \cos\iota  \right ] \;.
\end{eqnarray}
The parameterizations are consistent in the weak field, and are identical 
for zero spin. Rewriting Ryan's fluxes in terms of our parameters yields
\begin{eqnarray}
\dot{E} &=& -\frac{32}{5}\frac{\mu^2}{M^2} 
\left ( \frac{M}{p} \right )^{5} (1-e^2)^{3/2} 
\left [ f_{1}(e)  - \frac{a}{M} \left ( \frac{M}{p} \right )^{3/2} 
\cos\iota f_{2}(e) \right ]\;,
\label{Edot}
\\
\dot{L}_z &=& -\frac{32}{5}\frac{\mu^2}{M} 
\left (\frac{M}{p} \right )^{7/2}
(1-e^2)^{3/2} \left[ \cos\iota f_{3}(e) + \frac{a}{M} 
\left ( \frac{M}{p} \right )^{3/2}
\left[f_{4}(e) -\cos^2 \iota f_{5}(e)\right] \right]\;,
\label{Ldot} 
\\
\dot{C} &=& -\frac{64}{5} \mu^3 \left ( \frac{M}{p} \right )^{3}
(1-e^2)^{3/2}
\left [ f_{3}(e) -\frac{a}{M} \left (\frac{M}{p} \right )^{3/2}
\cos\iota f_{6}(e) \right ]\;,
\label{Qdot}
\end{eqnarray}
where $C\equiv Q + L_z^2$, and
\begin{eqnarray}
f_{1}(e) &=& 1+ \frac{73}{24}e^2 + \frac{37}{96} e^4\;,
\\
f_{2}(e) &=&  \frac{73}{12} + \frac{823}{24}e^2 + 
\frac{949}{32} e^4 + \frac{491}{192} e^6\;,
\\
f_{3}(e) &=&  1+ \frac{7}{8}e^2\;,
\\
f_{4}(e) &=& \frac{61}{24} + \frac{63}{8} e^2 + \frac{95}{64} e^4\;,
\\
f_{5}(e) &=& \frac{61}{8} + \frac{91}{4}e^2 + \frac{461}{64} e^4\;,
\\
f_{6}(e) &=& \frac{97}{12} + \frac{37}{2} e^2 + \frac{211}{32} e^4\;.
\end{eqnarray}  
In the $a=0$ limit, Eqs.\ (\ref{Edot}) and (\ref{Ldot}) reduce to the
celebrated Peters-Mathews formulae {\cite{pm}}.

The rates $\dot{q_j}=\{\dot{p},\dot{e},\dot{\iota}\}$ can be written
\begin{equation}
\dot{q}_j= H^{-1} ( b_j \dot{E} + c_j \dot{L_z} + d_j \dot{Q} )\;.
\label{rates}
\end{equation}
The quantities $H$ and $b_j$, $c_j$, $d_j$ are all constructed in a
straightforward way from derivatives of $E$, $L_z$, $Q$ with respect
to $p$, $e$, $\iota$; the resulting expressions are rather cumbersome,
and so are written out in Appendix~B.  We emphasize that these
functions encode the exact geodesic motion.

The main idea behind the hybrid scheme is simple: calculate the time
derivatives $\dot{q}_j$ using the {\em exact} coefficients $b_j$,
$c_j$, $d_j$ and the {\em approximate} fluxes (\ref{Edot}) --
(\ref{Qdot}).  A consistent leading-order calculation (that is,
leading order in $M/p$ and $a/M$) would instead approximate the
coefficients $q_j$ along with the fluxes.  Knowing the rates
$\dot{q}_j$ makes it possible to build the parameter space
trajectories $q_j(t)$ followed by a small body spiraling into a black
hole: given initial values $q_j(0)$, one simply ``integrates up'' the
derivatives $\dot{q}_j$ to generate the inspiral trajectory.  For
example, a simple-minded Euler-method integration would step from
parameter space coordinates $(t, q_j)$ to $(t + \delta t, q_j +
\dot{q}_j\delta t)$.  Generalization to more sophisticated integration
techniques is straightforward.  The trajectories $q_j(t)$ are the main
result of this paper.  From them, it is a simple matter to compute
quantities such as the gravitational waveform generated during an
inspiral, and thus to begin testing ideas more directly related to
data analysis.  We will not develop such waveforms here, but will
instead defer them to a later analysis.


\section{Comparison with numerical results}
\label{sec:compare}

The reliability of this method can be assessed by applying it to
specific families of orbits where accurate numerical results are
already known.  We first consider equatorial eccentric orbits,
recently studied by Glampedakis and Kennefick {\cite{kgdk}}.  Such
orbits always have $\iota = 0^\circ$ (prograde) or $\iota = 180^\circ$
(retrograde), leaving $p$ and $e$ as unspecified parameters.  Equation
(\ref{rates}) becomes
\begin{eqnarray}
\dot{p} &=& H_{\rm eq}^{-1} (-E_{,e} \dot{L}_z + L_{z,e} \dot{E} ) \;,
\\
\dot{e} &=& H_{\rm eq}^{-1} ( E_{,p} \dot{L}_z - L_{z,p} \dot{E} )\;,
\label{rates_eq}
\end{eqnarray}
where $H_{\rm eq} = E_{,p} L_{z,e} -E_{,e} L_{z,p}$.  The
leading-order approximation for these expressions is {\cite{cutler}}
\begin{eqnarray}
\dot{p} &=& -\frac{64}{5} \frac{\mu}{M} (1-e^2)^{3/2}
\left ( \frac{M}{p} \right )^{3} \left ( 1 + \frac{7}{8}e^2  \right )\;,
\label{pdot1}
\\
\nonumber \\
\dot{e} &=& -\frac{304}{15} \frac{\mu}{M^2} e(1-e^2)^{3/2}
\left ( \frac{M}{p} \right )^4 \left (  1 + \frac{121}{304} e^2
 \right )\;.
\label{edot1}
\end{eqnarray}
Note that we could equally well use the corresponding expressions with
the leading-order spin terms included (see \cite{kgdk}), but it turns
out that they essentially give the same results as Eqs.\ (\ref{pdot1})
and (\ref{edot1}).  These equations can be combined to give a simple
expression that describes the orbital evolution on the $p-e$ plane:
\begin{equation}
p(e) = p_i  \left ( \frac{e}{e_i}  \right)^{12/19} \left [ \frac{1 + 
121e^2/304}{1 + 121e_{i}^{2}/304 } 
\right ]^{870/2299}\;,
\label{peN}
\end{equation}
where $p_i$ and $e_i$ are initial values.  We are now ready to compare
the inspiral trajectories generated by Eq.\ (\ref{peN}) with those
obtained by the hybrid scheme.

Representative results for astrophysically relevant initial parameters
are shown in Fig.\ {\ref{fig1}}.  We compare the leading-order
trajectories found using Eq.\ (\ref{peN}) (dotted lines) with the
trajectories predicted by the hybrid scheme (solid lines).  Note that
the time dependence of the inspiral is suppressed in this figure: most
time is actually spent at large $p$.  The total duration of an
inspiral scales with $M^2/\mu$.  The shape of a curve, however, does
not depend on this ratio, provided that the mass ratio is extreme:
these curves are universal for $\mu\ll M$.  We show inspiral for both
prograde and retrograde orbits, for black hole spins $a = 0.5M$ and $a
= 0.9M$.

In all cases, the hybrid and the leading-order calculations agree for
$p\gg M$, as expected.  Differences between the two methods become
apparent in the strong field.  The leading-order inspiral trajectory
exhibits constantly decreasing eccentricity.  This is in marked
contrast with the rigorous strong-field calculations (numerical and
analytical) of Refs.\ {\cite{rr,cutler,kgdk}} showing that there
exists a region near the separatrix of stable/unstable orbits where
$\dot{e}$ reverses sign: the eccentricity should {\em grow} near the
separatrix.  It is very encouraging that the eccentricity does in fact
grow when the hybrid approximation is used.  Moreover, the location of
the critical points in these curves where $\dot{e} = 0$ is in good
agreement (at the order of a few percent) with the numerical results
of Refs.\ {\cite{cutler,kgdk}}; see Table {\ref{tab1}}.  Three of the
four cases shown in Fig.\ {\ref{fig1}} appear ``good'' in the sense
that the trajectories appear to agree reasonably well with what we
expect based on strong-field numerical analyses (cf.\ Ref.\
{\cite{kgdk}}).  The same comparison for the fourth case ($a = 0.9M$,
prograde; upper plot in the right hand panel of Fig.\ {\ref{fig1}})
reveals that both the eccentricity growth near the separatrix and the
distance of the critical curve $\dot{e}=0$ from the separatrix are
excessive. Prograde orbits of rapidly rotating black holes reach
rather deep into the black hole's strong field where the weak-field
fluxes (\ref{Edot}) and (\ref{Ldot}) cannot be trusted.  As we shall
see when we move on to generic inspirals (Sec.\ {\ref{sec:generic}}),
this breakdown of the weak-field flux formulae means that the hybrid
approximation does not accurately describe the inspiral of shallow
inclination orbits ($\iota\lesssim 20^\circ$ or so) into rapidly
rotating black holes ($a\gtrsim 0.85M$ or so).

\begin{figure}[tbh]
\centerline{
\epsfxsize=9cm \epsfbox{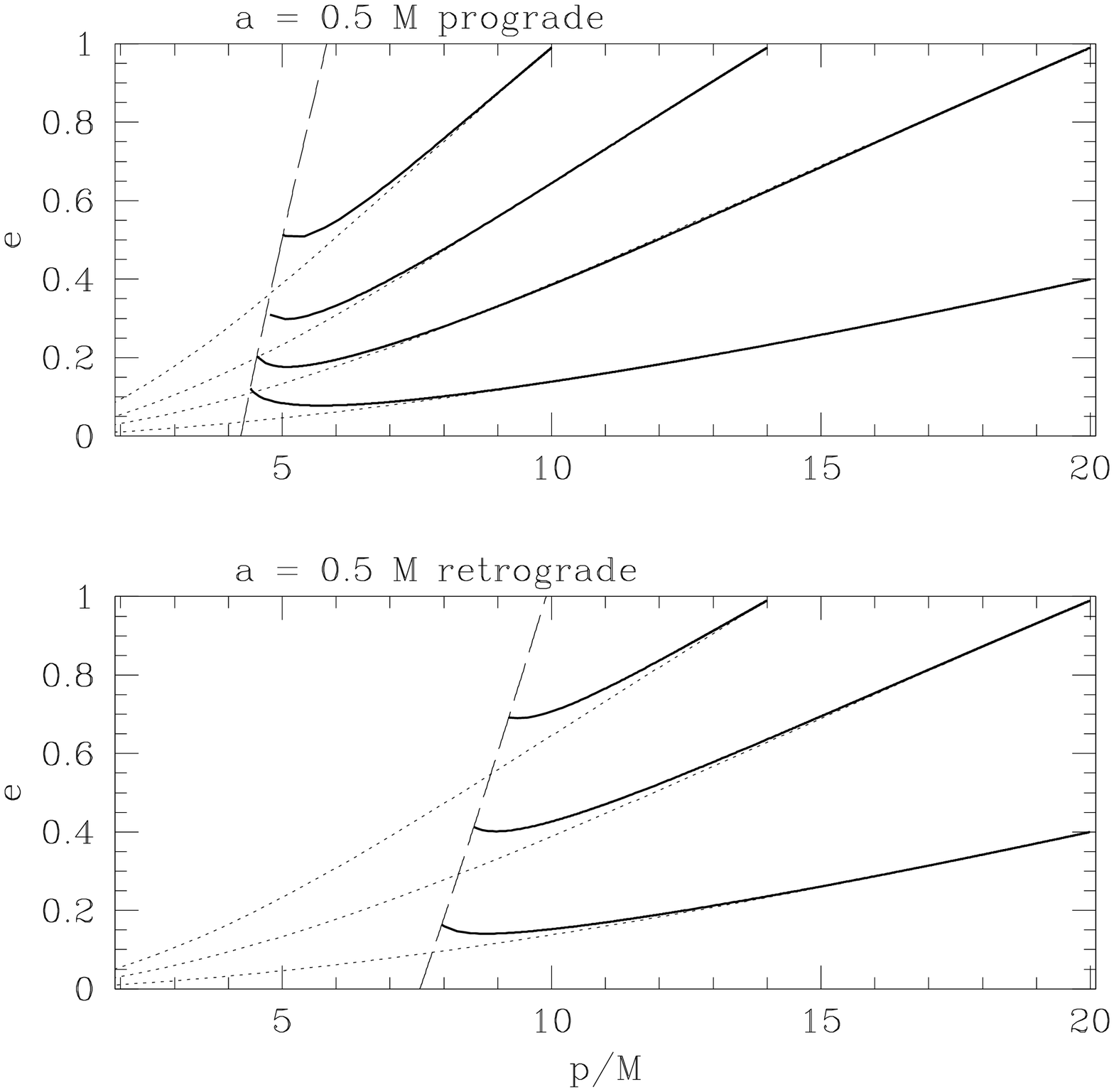}
\epsfxsize=9cm \epsfbox{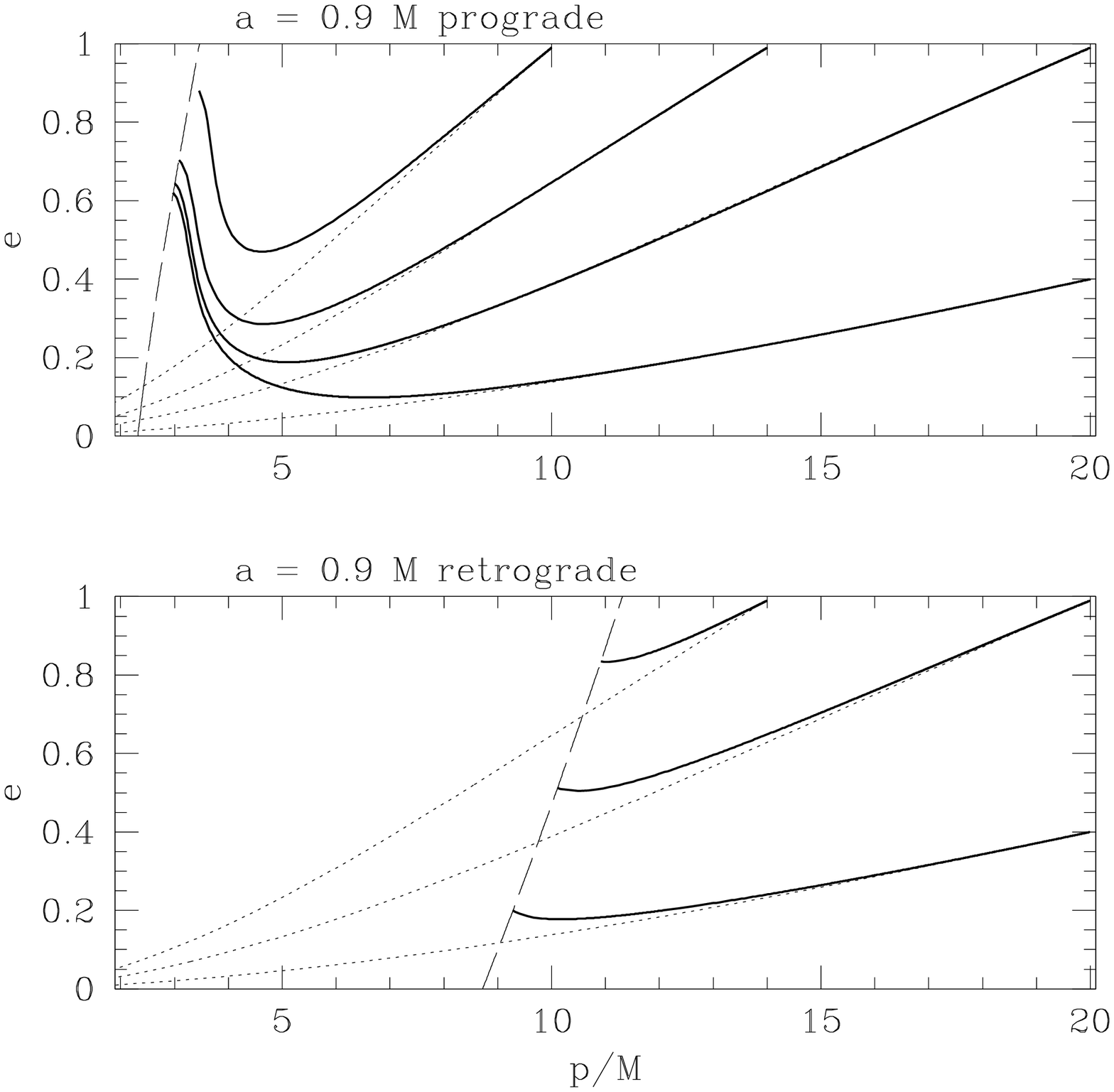}
}
\vspace{0.1cm}
\caption{Comparing equatorial inspiral.  We show inspiral into a hole
with spin $a = 0.5 M$ (left panel) and into a hole with $a = 0.9M$
(right panel).  In each panel, the top half shows prograde inspirals
and the bottom retrograde inspirals.  In each set, the dashed line
represents the separatrix separating stable from unstable orbits.  We
used the hybrid approximation discussed in the text to radiatively
evolve orbits with initial parameters $(p_i, e_i) = (20M, 0.4)$,
$(20M, 0.99)$, $(14M, 0.99)$, and $(10M, 0.99)$.  The inspiral
trajectories are shown as the heavy lines in each plot.  (The final
set is not included in the retrograde inspirals since the initial
conditions are not stable in those cases.)  The dotted trajectories in
each plot shows the leading-order predictions generated using Eq.\
(\ref{peN}).  Note the significant qualitative difference between the
two calculations at the vicinity of each separatrix.  Note also the
extremely large growth in eccentricity seen in the prograde inspirals
for $a = 0.9M$ just before reaching the separatrix. Comparison with
accurate strong-field numerical results shows that this growth is 
excessive.}
\label{fig1}
\end{figure}

It is possible to get some insight into the superior qualitative
description of the inspiral in the strong field region given by our
approach. The phenomenon of orbital circularization as a result of
some form of dissipation is seen in many astrophysical situations,
such as that of satellites whose orbits are decaying due to
atmospheric friction. The reason is that the dissipating mechanism
causes the particle to ``drop'' in its potential well, the usual
geometry of which ensures that the orbital eccentricity decreases. In
our case another mechanism becomes significant as the unstable plunge
orbit is approached at the end of the inspiral. As this occurs the
potential becomes shallower (as the minimum turns into a saddle point
at plunge), and this tends to increase the eccentricity of the
orbit. Shortly before plunge this mechanism overcomes the
circularizing tendency. It is not surprising that the hybrid
approximation can qualitatively replicate the eccentricity increasing
behavior, because it exactly describes the shape of the orbital
potential, which is so critical to this effect.

Table {\ref{tab2}} compares data for $\dot{p}$ and $\dot{e}$.  In this
sample, the hybrid approach clearly is more accurate than the
leading-order approximation.  This comparison is a very strict test of
the accuracy of this scheme.  As discussed above, we believe that the
hybrid approximation is reliable as long as $r_p\gtrsim 5M$.  The
weak-field fluxes that we use cannot be trusted very deep in the
strong field --- the spin correction terms in Eqs.\ (\ref{Edot}) and
(\ref{Ldot}) dominate the leading order term.  The method therefore
fails when we push to smaller $r_p$.  This effectively constrains the
black hole spin to $a\lesssim 0.5M$ for prograde motion --- for larger
spins, the innermost stable orbit and hence $r_p$ come too close to
the horizon.  For retrograde orbits, the results are much more
accurate since $r_p$ never comes close to the horizon, regardless of
the spin.  Finally, we emphasize the essential role Ryan's fluxes
(\ref{Edot}) and (\ref{Ldot}) play in calculating $\dot{p}$ and
$\dot{e}$.  Had we used instead the Peters-Mathews fluxes, the
resulting inspirals would predict a rapid circularization under
radiation reaction: we find that the Peters-Mathews fluxes reduce the
eccentricity to zero well before reaching the saddle point of the
orbital potential, and so the eccentricity never grows.  This is in
sharp disagreement with the numerical results.

A major prediction of the hybrid approximation is that for equatorial
orbits the residual eccentricity prior to plunge should be
substantial, in strong contrast to the prediction of the leading order
formula (\ref{peN}).  In many cases, the leading order results predict
that the orbit will actually circularize prior to plunge.  Because the
harmonic structure of a circular inspiral is rather different from
that of an inspiral with substantial eccentricity, these results have
strong implications for the waveform models to be used in LISA's data
analysis.

We next consider circular inclined orbits, which were recently studied
by Hughes {\cite{scott1}}.  One of the most important findings of
Ref.\ {\cite{scott1}} is that the angle $\iota$ remains almost
constant during inspiral, even when the particle is crossing strong
field regions.

For these orbits, the rates $\dot{p}$ and $\dot{\iota}$ are given by
\begin{eqnarray}
\dot{p} &=& H_{\rm circ}^{-1} ( -L_{z,\iota} \dot{E} + E_{,\iota}
\dot{L}_z )\;,
\\
\dot{\iota} &=& H_{\rm circ}^{-1} ( L_{z,p} \dot{E} - E_{,p}
\dot{L}_z )\;,
\label{rates_c}
\end{eqnarray}
where $H_{\rm circ} = E_{,\iota} L_{z,p} - L_{z,\iota} E_{,p}$.  In
order to obtain these formulae we first expressed $\dot{Q}$ in terms
of $\dot{E}$ and $\dot{L}_z$ making use of the ``circular goes to
circular'' theorems {\cite{ryan1,ori,minothesis}}; see Ref.\
{\cite{scott2}} for further discussion.  The leading-order expression
for $\dot{\iota}$ is {\cite{ryan2}}
\begin{equation}
\dot{\iota}= \frac{244}{15} \frac{\mu}{M^2} \frac{a}{M} 
\left ( \frac{M}{p} \right )^{11/2} \sin\iota \;;
\label{idotN}
\end{equation}
$\dot{p}$ follows from Eq.\ (\ref{pdot1}), setting $e=0$.

\begin{figure}[tbh]
\centerline{
\epsfxsize=10cm \epsfbox{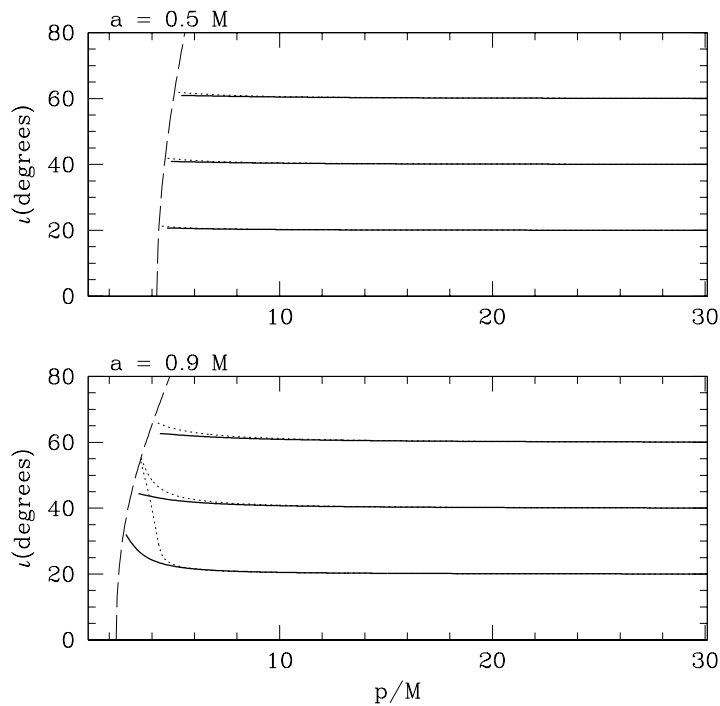}
}
\vspace{0.1cm}
\caption{Comparing circular, inclined inspiral.  We show inspiral
into a hole with spin $a = 0.5M$ (top) and spin $a = 0.9M$ (bottom).
The solid lines show inspiral using the hybrid approximation; the
dotted lines show the leading order inspiral prediction.  The dashed
curve shows the separatrix between stable and unstable orbits.  Both
approximations show that the inclination angle increases, especially
right before reaching the separatrix.  However, the increase predicted
by the leading-order prediction is far too large, particularly for
rapidly spinning black holes.  The inspiral predicted by the hybrid
approximation is closer to what is seen in rigorously computed
inspirals.  Nonetheless, it too shows an increase in $\iota$ that is
probably excessive.  As we argue in the text, holding $\iota$ constant
produces an inspiral sequence that is probably closest of all to
strong-field calculations and should be acceptably accurate.}
\label{fig2}
\end{figure}

Table {\ref{tab3}} compares data for $\dot{p}$ and $\dot{\iota}$ using
the hybrid approximation to the the results obtained using Eqs.\
(\ref{pdot1}) and (\ref{idotN}), together with accurate numerical
results from Ref.\ {\cite{scott1}}.  Figure {\ref{fig2}} shows
inspirals of circular inclined orbits with our method and using the
leading-order formulae.  Both approximations predict that $\iota$
changes in such a way as to drive the orbit to an equatorial
retrograde configuration (that is, $\iota$ increases).  The two
calculations agree at large radii.  In the strong field, the
leading-order formulae break down --- the inclination angle tends to
increase dramatically.  The behavior of the hybrid-scheme inspiral is
more reasonable.

Although the hybrid scheme is much better behaved in the strong field,
the growth of $\iota$ we see is still quite a bit larger than detailed
numerical calculations predict {\cite{scott1,scott2}}.  Based on those
numerical results, a more accurate scheme would be to simply require
that $\iota$ remain constant.  Combining $d\iota/dt = 0$ with Eq.\
(\ref{ioteq}) yields the rule
\begin{equation}
\dot{Q} = \frac{2Q}{L_z} \dot{L}_z\;.
\label{Qdotc}
\end{equation}
This rule is consistent with our description of $Q$ as roughly the
squared component of angular momentum projected into the equatorial
plane.  If the spacetime is perfectly spherical (i.e., Schwarzschild
black holes), then $Q$ is exactly such an angular momentum component:
$Q_{\rm spherical}\equiv L_x^2 + L_y^2$. It is easy to show that an
inspiral in this spacetime would proceed at exactly constant
inclination angle: gravitational waves carry off exactly the right
amounts of $L_x$ and $L_y$ to hold $\iota$ constant, so Eq.\
(\ref{Qdotc}) is exactly correct in this case.  One would expect
$\iota$ to remain nearly constant if the spacetime does not deviate
too strongly from sphericity.  Rigorous numerical results for the
circular inclined case show that $\iota$ indeed remains nearly
constant; it thus appears that the Kerr metric is not too aspherical
over much of the inspiral (modulo frame dragging).  Additional
evidence for the validity of this statement is given by the discussion
in Appendix {\ref{app:almostsphere}}.  Since the orbit's eccentricity
does not enter this argument at all, it is likely that Eq.\
(\ref{Qdotc}) will work well for inclined circular orbits also.


\section{Evolving generic orbits}
\label{sec:generic}

Having established the reliability and limitations of the hybrid
scheme, we move to the main subject of this paper: the study of
inspirals of test bodies in generic orbits where only leading-order
results are currently available \cite{ryan1}.

We began this analysis employing Ryan's fluxes, Eqs.\ (\ref{Edot}) --
(\ref{Qdot}), but quickly faced disappointing results.  We found that
hybrid-scheme inspirals produced with these fluxes did not behave well
far from the two limits discussed above, particularly in the strong
field.  For example, the eccentricity tended to grow extremely large
very rapidly in some cases.  The root of the problem lies in the
expression for the $\dot{Q}$ flux, Eq.\ (\ref{Qdot}), which apparently
is not as accurate as we would require it to be.  The qualitative
behavior of our inspirals is more reasonable when the rule given by
Eq.\ (\ref{Qdotc}) is used to compute $\dot Q$ instead, forcing
$\iota$ to be constant.  Following the discussion at the end of
Section {\ref{sec:compare}} and in Appendix {\ref{app:almostsphere}},
it is likely that this rule is accurate enough for our purposes
anyway, and so we shall use it from this point onward.  In all
likelihood, detailed self-force calculations will be needed to test
the accuracy of the constant inclination rule.

Using the fluxes given in Eqs.\ (\ref{Edot}) and (\ref{Ldot}) with the
constant-$\iota$ rule (\ref{Qdotc}) produces inspirals that agree with
the leading-order results when $p\gg M$, that smoothly converge to the
equatorial case for $\iota\to 0^\circ$ and $\iota\to 180^\circ$, and
that exhibit an $\dot{e} > 0$ region near the separatrix.  Examples of
this behavior are shown in Fig.\ {\ref{fig3}} (spin $a = 0.5M$) and
Fig.\ {\ref{fig4}} (spin $a = 0.9M$).  Several interesting features
can be seen in these figures.  The trajectories for $\iota < 90^\circ$
are qualitatively similar to the equatorial, prograde trajectories
shown in Fig.\ {\ref{fig1}}.  In particular, each such trajectory
passes through a critical point at which $\dot e = 0$ after which
eccentricity grows.  The system typically has substantial non-zero
eccentricity when it reaches the separatrix.  Also, note the excessive
growth in eccentricity near the separatrix for $a = 0.9M$ and $\iota =
30^\circ$.  At shallow inclination angle, the separatrix is very deep
in the black hole's strong field, so the inspiral proceeds to small
$r$ before plunging.  Just as in the case of equatorial orbits for $a
= 0.9M$, the weak-field flux formulae that we use cannot be trusted
this far into the Kerr black hole strong field.

The qualitative appearance of the inspirals for $\iota > 90^\circ$ is
quite a bit different from the $\iota < 90^\circ$ inspirals.  Looking
at the right hand sides of Figs.\ {\ref{fig3}} and {\ref{fig4}}, we
see that the hybrid approximation predicts that many of these
inspirals completely circularize prior to reaching the separatrix.  We
do not believe that this behavior is robust.  Indeed, we find that the
behavior of inspirals exhibits a rather sharp transition as the
inclination angle goes from slightly below $90^\circ$ to slightly
above.  This behavior arises from the $\cos\iota$ terms in Eqs.\
(\ref{Edot}) and (\ref{Ldot}), which switch sign at this transition.
We thus do not believe that this rapid circularization is physical,
but instead attribute it to poor behavior of the hybrid approximation
at $\iota\gtrsim 90^\circ$.

\begin{figure}[tbh]
\centerline{
\epsfxsize=9cm\epsfbox{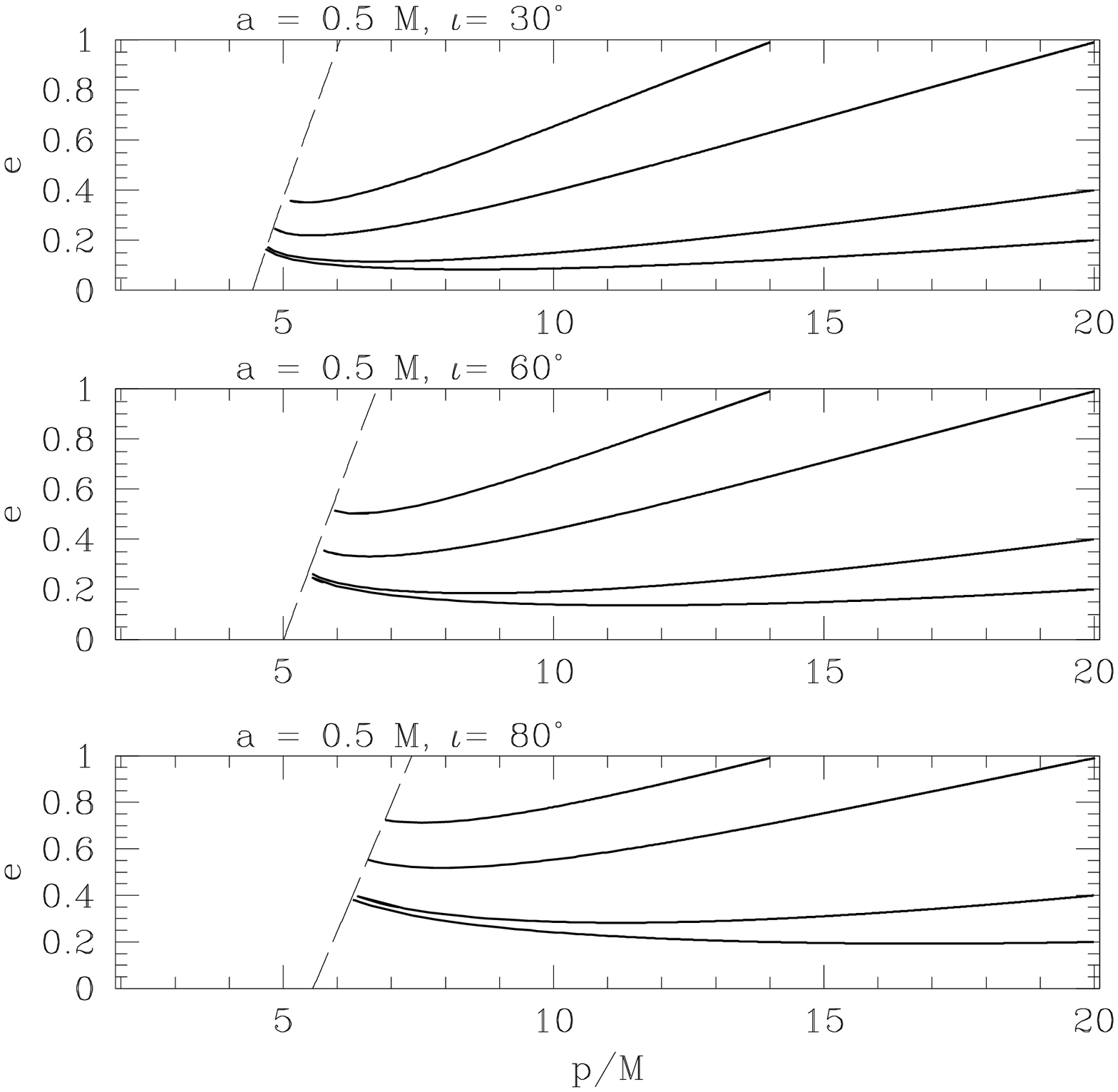}
\epsfxsize=9cm\epsfbox{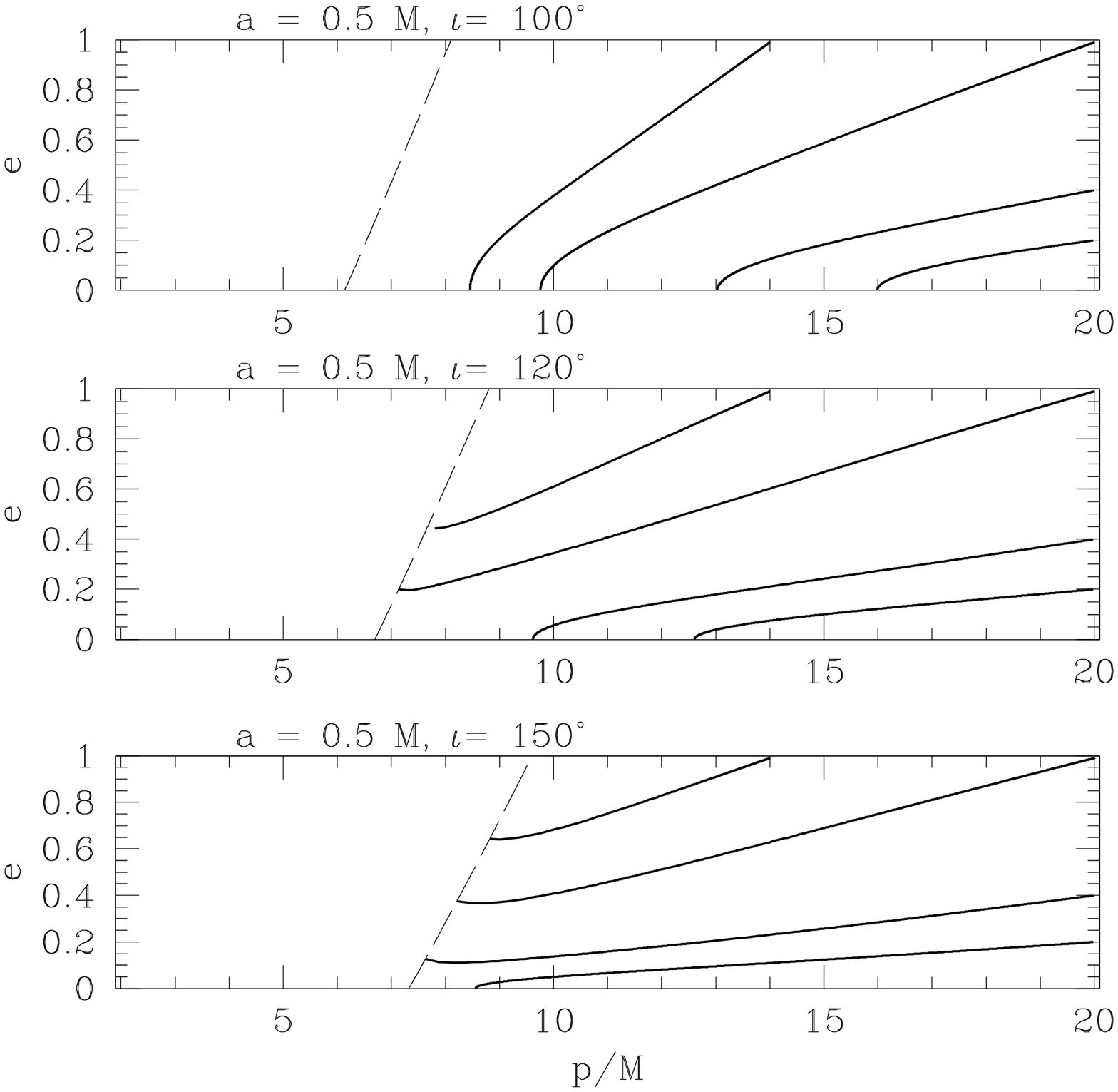}
}
\vspace{0.1cm}
\caption{Comparing generic inspiral at several inclination angles
into a hole with spin $a = 0.5M$.  In all plots, the dashed line
represents the separatrix between stable and unstable orbits; because
we force inspiral to lie in surfaces of constant $\iota$, there exists
a separatrix in the $p-e$ plane of each such surface.  We use the
hybrid inspiral scheme to evolve orbits with initial parameters $(p_i,
e_i) = (20M, 0.2)$, $(20M, 0.4)$, $(20M, 0.99)$, and $(14M, 0.99)$,
and $\iota = 30^\circ$, $60^\circ$, $80^\circ$, $100^\circ$,
$120^\circ$, and $150^\circ$.  Inspirals that are roughly ``prograde''
(have $\iota < 90^\circ$) are qualitatively similar to the equatorial
inspirals shown in Fig.\ {\ref{fig1}}: there is a turnaround in the
eccentricity evolution, so that all inspirals finish with a
substantial non-zero eccentricity.  By contrast, the roughly
``retrograde'' inspirals ($\iota > 90^\circ$) exhibit rather different
behavior: particularly when the inclination is not too far from
$90^\circ$, many inspirals completely circularize, reaching $e = 0$.
As discussed in the text, we believe this behavior is incorrect.}
\label{fig3}
\end{figure}

Having established that the behavior of hybrid approximation inspirals
for $\iota\gtrsim 90^\circ$ is probably not reliable, it is worth
re-examining the behavior for $\iota\lesssim 90^\circ$.  Good examples
of this behavior are the plots for $\iota = 80^\circ$ (lower leftmost
panels of Figs.\ {\ref{fig3}} and {\ref{fig4}}).  In these cases, the
$\cos\iota$ terms in Eqs.\ (\ref{Edot}) and (\ref{Ldot}) will be small
but positive.  Indeed, we see that the trajectories shown in this case
are somewhat odd.  Consider the trajectories that begin at $(p_i, e_i)
= (20 M, 0.2)$.  The points where the eccentricity evolution switches
sign are at rather large semi-latus rectum ($p\sim 16.5 M$ for $a =
0.5M$; $p\sim 25M$ for $a = 0.9M$).  This is quite a bit further out
than is seen in any analysis of radiation reaction on equatorial
orbits {\cite{rr,cutler,kgdk}}.  We suspect that this behavior is
likewise an artifact of the weak-field fluxes, and do not trust the
hybrid approximation's predictions for inspirals near $\iota =
90^\circ$.

We conclude that the hybrid inspiral scheme --- the weak-field fluxes
(\ref{Edot}) and (\ref{Ldot}) plus the ``constant inclination'' rule
(\ref{Qdotc}) applied to exact, strong-field Kerr geodesics --- is, in
most cases, reliable and accurate enough to be used for exploring
issues in LISA's data analysis.  In some cases, the hybrid scheme will
{\it not} be accurate enough: the weak-field fluxes behave badly when
the orbit goes too deep into the strong field, and so we do not trust
this scheme's results when $r_p\lesssim 5M$.  Also, the spin
correction terms in Eqs.\ (\ref{Edot}) and (\ref{Ldot}) become either
very small or very large relative to the leading quadrupole term when
$\iota\sim90^\circ$, and so we do not trust the hybrid approximation
for inclination angles $80^\circ\lesssim\iota\lesssim120^\circ$.  More
rigorous strong-field analyses will be needed in order to validate the
inspiral behavior at these inclination angles.

In all cases in which the inspiral behavior is reasonable, we find
that small body's orbits will have significant eccentricity upon
reaching the separatrix.  Eccentricity will be a significant factor
that must be incorporated into plans for LISA data analysis.

\begin{figure}[tbh]
\centerline{
\epsfxsize=9cm \epsfbox{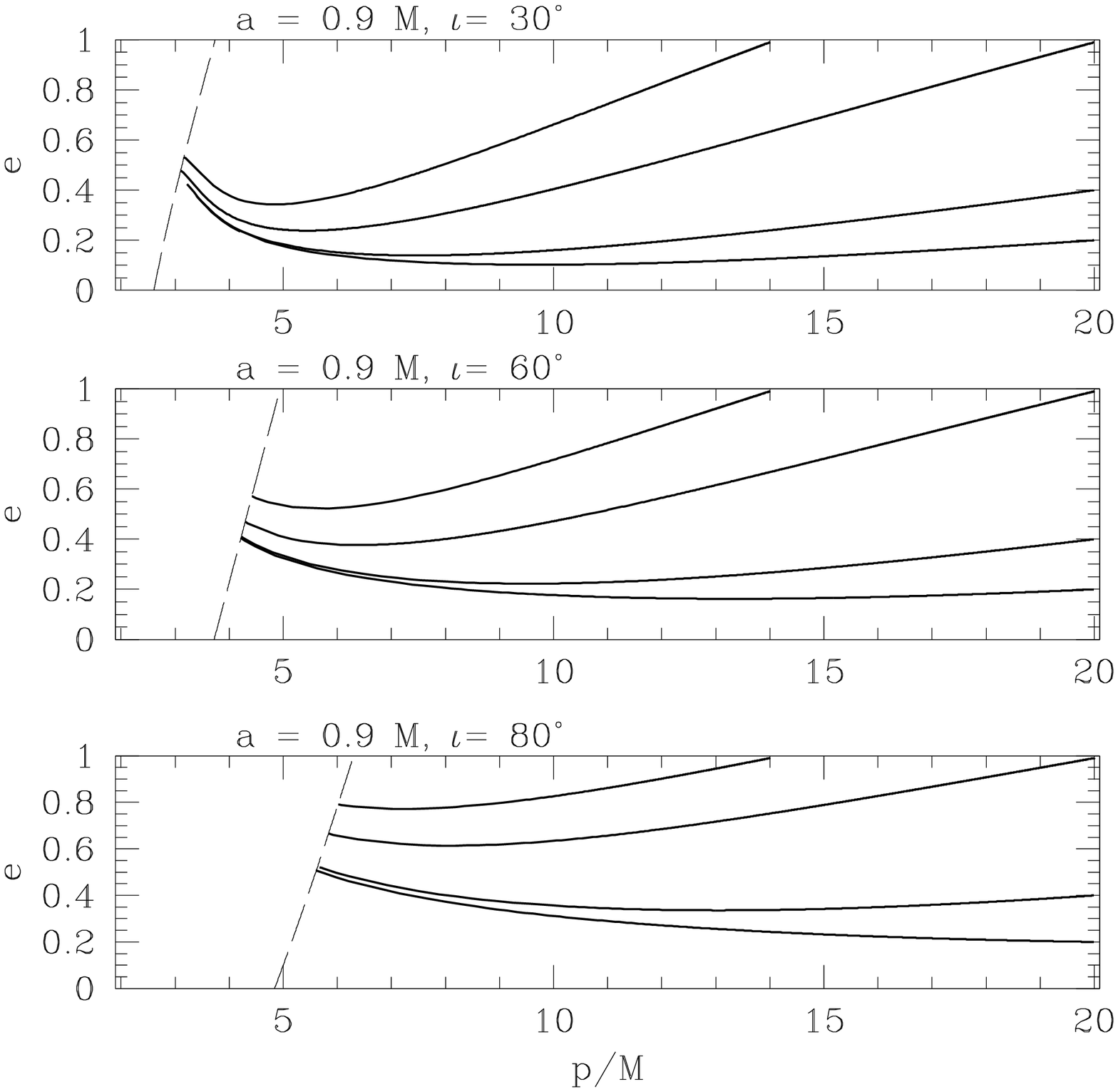}
\epsfxsize=9cm \epsfbox{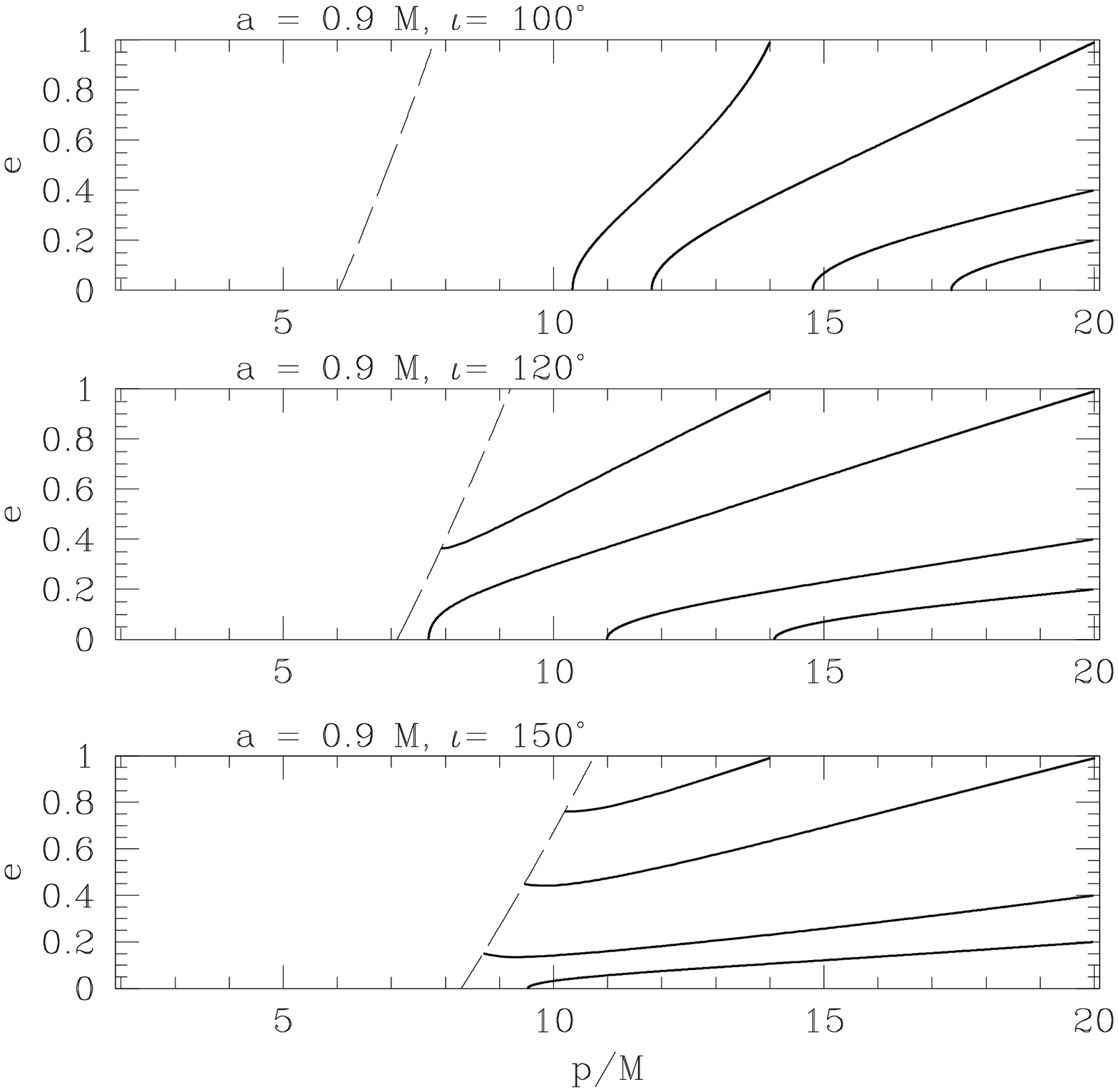}
}
\vspace{0.1cm}
\caption{Comparing generic inspiral at several inclination angles
into a hole with spin $a = 0.9M$.  Aside from the faster black hole
spin, the trajectories shown here have identical initial parameters as
those shown in Fig.\ {\ref{fig3}}.  The inspirals for $\iota <
90^\circ$ are again qualitatively similar to equatorial inspirals,
shown in Fig.\ {\ref{fig2}}.  In particular, we see that at shallow
inclination angle, the growth in eccentricity is very large.  We
believe this is because these orbits go so deeply into the
strong-field that the weak-field flux formulae used in the hybrid
approximation are no longer reliable.  We also see the rapid
circularization of inspirals when $\iota$ is greater than but near
$90^\circ$, very similar to the behavior encountered for spin $a = 0.5
M$.}
\label{fig4}
\end{figure}
 

\section{Conjecture: approximate $\dot{Q}$ for generic zoom-whirl orbits}
\label{sec:zoomwhirl}

We have repeatedly emphasized that the hybrid approximation presented
here is reliable as long as the orbiting body does not come too close
to the central black hole.  This excludes shallow inclination orbits
of rapidly rotating black holes --- an unfortunate exclusion, since
those orbits are in many cases very strongly ``stamped'' by the
features of the strong-field spacetime, and thus may be the most
interesting inspiral sources observed by LISA {\cite{nhs}}.  These are
also the orbits for which the ``constant inclination $\dot Q$'' rule
(\ref{Qdotc}) is most likely to be inaccurate, since they are deepest
in the Kerr black hole's strong field and are most likely to feel the
asphericity of the Kerr spacetime.  Ultimately, self-force
computations will provide the toolsets needed to rigorously compute
strong-field inspirals in this regime, and will sidestep all
difficulties regarding the calculation of $\dot Q$.  In the meantime,
while such computations remain unavailable, it is worth exploring
other possibilities that may provide accurate approximations to the
Carter constant's evolution.

Strong-field eccentric equatorial orbits of rapidly rotating holes
have a ``zoom-whirl'' character {\cite{kgdk}}: orbits near the
separatrix in the $p-e$ plane ``whirl'' around the black hole
repeatedly near periastron, so that the amount of azimuth $\phi$
accumulated in a single radial cycle (apastron to periastron to
apastron) is much greater than $2\pi$.  During this whirl phase, the
body's motion is very close to a circular orbit.  Exploratory studies
{\cite{tdc_sah}} show that this behavior carries over to
non-equatorial orbits, particularly for fairly shallow inclination
angle.

The equatorial zoom-whirl orbits studied in Ref.\ {\cite{kgdk}}
radiate energy and angular momentum as if they were nearly circular:
\begin{equation}
\dot{E} \approx \Omega_{\phi} \dot{L}_{z} \;,
\label{zw_eq}
\end{equation}
where $\Omega_{\phi}$ is the frequency associated with azimuthal
motion.  This property follows quite naturally from the motion of a
test-body in a zoom-whirl orbit: a large fraction of the orbital
period is spent ``whirling'' in the vicinity of the periastron, where
the motion is nearly circular.  This is also the part of the orbit
where the body is closest to the black hole and thus where most of the
radiation will be generated.  Thus, the radiation from a zoom-whirl
orbit should be very similar to radiation from a circular orbit, which
is exactly what Eq.\ (\ref{zw_eq}) suggests.  Extrapolating this
behavior to generic zoom-whirl orbits, we expect that most of the
radiated $E$, $L_z$, and $Q$ will come from the motion of the body
near a (generalized) separatrix in the $(p,\iota,e)$ phase space.  The
whirl motion of such orbits will be well-described as nearly circular
and inclined.

Following Kennefick and Ori {\cite{ori}}, we can write the Carter
constant as
\begin{equation}
Q = G(r,E,L_z) -\Delta u_{r}^{2} \;,
\end{equation}
where
\begin{equation}
G = \Delta^{-1} [ E (r^2+a^2) -aL_z ]^2 -(L_z -aE)^2 -r^2 \;,
\end{equation}
with $\Delta = r^2 - 2Mr + a^2$ and $u_r$ denoting the radial
component of the body's four-velocity.  (The function that we have
denoted $G$ is written $H$ in {\cite{ori}}.)  It is then
straightforward to show that {\cite{ori}}
\begin{equation}
\dot{Q}= G_{,E} \dot{E} + G_{,L_z} \dot{L}_z -
\frac{2\Sigma u^{r}}{u^{t}} F_{r} \; ,
\label{Qdot_zw}
\end{equation} 
where $\Sigma = r^2 + a^2\cos^2\theta$ and $F_{r}$ is the radial
component of the self force.  It is the unknown last term in this
equation that presently prohibits the calculation of $\dot{Q}$ for
generic orbits.  For strictly circular orbits, on the other hand, this
term is absent since $u^r=0$.  The remaining expression $\dot Q =\dot
Q(\dot{E},\dot{L}_z)$ allows the explicit calculation of $\dot{Q}$;
this is how Hughes evolves circular, inclined orbits by reading the
fluxes $\dot{E}$ and $\dot{L}_z$ at infinity and down the hole
{\cite{scott1,scott2}}.

For a zoom-whirl orbit and for motion near the periastron, $r\approx
r_p$, so we should have $u^{r} \approx 0$; consequently, the unknown
term in (\ref{Qdot_zw}) should be negligible.  Our {\em conjecture} is
that the resulting expression for $\dot{Q}$ describes the evolution of
the Carter constant for all generic zoom-whirl orbits and with
increasing accuracy as the orbit approaches the separatrix.  We
emphasize that this approximation should hold even for orbits deep in
the black hole's strong-field.  This conjecture could become a
practical tool once a code that calculates $\dot{E}$ and $\dot{L}_z$
for generic orbits is developed.  Furthermore, a direct comparison
between (\ref{Qdotc}) and (\ref{Qdot_zw}) should be a useful guide for
the accuracy of the $\iota=\mbox{constant}$ rule in strong-field
situations.  Future computation of the self force will provide the
ultimate test for both approximations.


\section{Concluding discussion}
\label{sec:conclude}

The hybrid approximation presented in this paper should be a valuable
tool for the gravitational-wave astrophysics community as studies of
LISA's data analysis requirements begin, and thence models of the
waves generated by compact bodies spiraling into massive black holes
become needed.  Such approximate ``kludged'' waveforms are obviously
too crude to actually be used in future measurements of compact bodies
spiraling into massive black holes; data analysis strategies based on
waveforms built from rigorous strong-field radiation reaction will be
needed.  Waveforms from approximate inspiral models should be adequate
to begin the process of developing a data analysis infrastructure.
For example, they will be useful for counting the number of analysis
filters needed, assessing the computational cost of data analysis, and
experimenting with data analysis techniques.  As rigorous and reliable
waveform models become available, they can simply be dropped into the
codes and infrastructure developed using the hybrid approximation.

Because this approximation combines the exact strong-field Kerr
geodesics with weak-field radiation reaction formulae, it is somewhat
limited: inspiral cannot go too deeply into the strong field, thereby
making it inaccurate for shallow ($\iota\lesssim20^\circ$) inspirals
of rapidly rotating ($a\gtrsim0.85M$) holes.  Also, the $\cos\iota$
dependence of terms within the flux formulae behaves badly near
$\iota\sim90^\circ$, so that the approximation is probably not
reliable within an inclination range $80^\circ\lesssim\iota\lesssim
120^\circ$.  Despite these limitations, we have found the hybrid
approximation reliably and robustly duplicates many of the inspiral
properties that we expect will be found when it is possible to study
these systems using truly strong-field gravitational radiation
reaction.  In particular, it produces inspiral trajectories that
retain substantial non-zero eccentricity just before plunging into the
hole, as is expected from strong-field analyses in special cases
{\cite{rr,cutler,kgdk}}.  We emphasize this point because the harmonic
structure of gravitational waves from eccentric orbits is quite a bit
different from that of waves generated by circular orbits.  The
residual eccentricity of typical inspirals is likely to impact data
analysis rather strongly.

Obviously, waveforms constructed from hybrid approximation inspirals
are by no means the ultimate models that will be needed for LISA data
analysis --- we strongly advocate continuing to develop techniques for
understanding strong-field radiation reaction.  Future insight from
such studies may make it possible to improve the hybrid approximation.
Even when strong-field radiation reaction is mature enough to model
arbitrary compact body inspirals, the calculation may be
computationally expensive, so that an approximation scheme of some
sort may remain useful.

Although our overall goal is to provide a tool that can be used to
model the gravitational waves produced by compact body inspiral, we
have presented no such waves in this paper.  That is the next step.
The calculations we have presented explicitly construct the parameter
space trajectories $[E(t), L_z(t), Q(t)]$ describing an inspiral.  It
is then a simple matter to combine such a trajectory with the geodesic
equations for the Kerr spacetime {\cite{mtw}} to produce the
Boyer-Lindquist coordinate space inspiral $[r(t),\theta(t),\phi(t)]$.
This set of functions specifies the worldline of the inspiraling body,
and one can use it to compute the gravitational waveform seen by
distant observers (see, for example, Ref.\ {\cite{press77}}).  Codes
to perform this next step are under development {\cite{tdc_sah}};
results should be presented in the near future.


\acknowledgements

We thank Kip Thorne for pressing us to develop ``fast and dirty''
techniques to compute inspiral waveforms, and Teviet Creighton for
helping to test and debug the code that underlies parts of this
analysis.  K.\ G.\ thanks B.\ S.\ Sathyaprakash and Nils Andersson for
useful interactions related to this work and also acknowledges support
from PPARC Grant PPA/G/0/1999/0214.  S.\ A.\ H.\ is supported by NSF
Grant PHY-9907949.  D.\ K.\ is partially supported by NSF Grant
PHY-0099568.


\appendix

\section{}
\label{app:almostsphere}

The occurrence of a ``third'' orbital constant $Q$ in axisymmetric
gravitational fields is not an exclusive feature of general
relativity.  For example, it is familiar from Newtonian celestial
mechanics applied to orbital motion in galactic gravitational
potentials (see, for example Ref.\ {\cite{galactic}}, where the third
constant is denoted $I$).  The departure of $Q$ from $L_x^2 + L_y^2$
can then be attributed to the asphericity of the potential.  If such a
potential does not deviate very much from sphericity, $L^2$ (the
square of the total angular momentum) turns out to be almost constant,
so that $Q$ should be, after all, nearly $L^2 - L_z^2$.

It is straightforward to check whether this behavior of $L^2$ occurs
in Kerr spacetime.  The definition we use for $L^2$ is identical to
that used in Schwarzschild spacetime,
\begin{equation}
L^2 = p_{\theta}^2 + (\sin\theta)^{-2} p_{\phi}^2\;,
\label{Ltot}
\end{equation}
where $p_{a}$ denotes the particle's four-momentum.  For the Carter
constant $Q$ we have {\cite{mtw}}
\begin{equation}
Q = p_{\theta}^2 + (\sin\theta)^{-2} p_{\phi}^2 -p_{\phi}^2 +
a^2\cos^2\theta(1 - E^2)\;.
\label{Qeq}
\end{equation}
Combining these two expressions gives
\begin{equation}
Q = L^2 - L_{z}^2 + a^2 \cos^2\theta(1 - E^2)\;.
\label{QnL}
\end{equation}
In other words, $Q$ can be interpreted as the projection of the total
angular momentum on the equatorial plane, modulo the ``aspherical''
term $a^2\cos^2\theta(1 - E^2)$.  This interpretation makes sense when
the aspherical term is small --- that is, when $a\ll M$ (slow
rotation) and/or $E\approx 1$ (weak-field orbits).  In practice, we
find that this term is often significantly smaller than the preceding
terms even for motion in strong-field regions of rapidly rotating
holes.  We illustrate this in Fig.\ {\ref{fig5}}, showing how the
quantity $\delta L^2 \equiv L^2/(Q + L_z^2) - 1$ varies with time for
a variety of generic orbits around a rapidly spinning hole.

Examining Fig.\ {\ref{fig5}}, we see that $L^2$ deviates very little
from $Q + L_z^2$ even when the small body is deep in the black hole's
strong field --- in this sample, the difference is no more than about
$1\%$.  This shows that interpreting $Q$ as a squared projection of
angular momentum into the equatorial plane is sensible.  Because $Q +
L_z^2$ is a constant quantity, this figure also demonstrates that
$L^2$ is nearly constant.  This is exactly what we expect for motion
in an axisymmetric potential that is almost spherical.  These pieces
of evidence suggest that the Kerr spacetime is not as ``aspherical''
as we might have expected, at least for the purposes of this argument,
lending credence to our suggestion that the ``$\iota
=\mbox{constant}$'' assumption should be reliable, as discussed in the
paper's main body.

\begin{figure}[tbh]
\centerline{
\epsfxsize=12cm\epsfbox{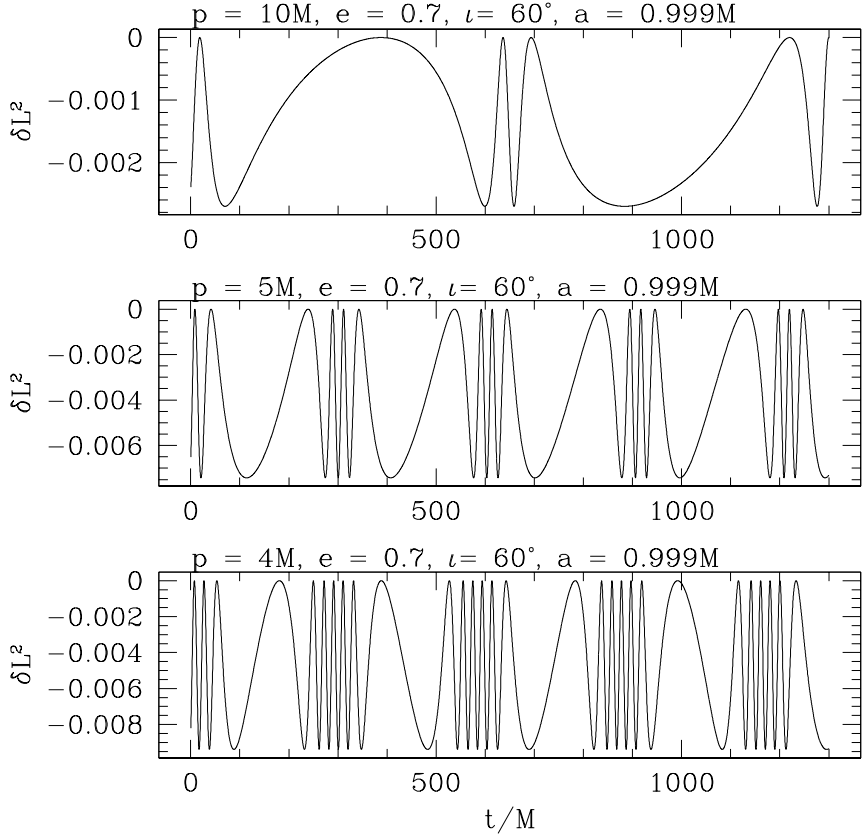}
}
\vspace{0.1cm}
\caption{Examining our notion of ``total angular momentum'' for
strong-field Kerr black hole orbits.  Each panel compares the angular
momentum squared $L^2\equiv p_\theta^2 + (\sin\theta)^{-2} p_\phi^2$
to $Q + L_z^2$: the quantity plotted is $\delta L^2\equiv L^2/(Q +
L_z^2) - 1$.  The top panel shows these quantities over an orbit with
$p = 10M$, the center panel the quantities over an orbit with $p =
5M$, and the bottom over an orbit $p = 4 M$.  In all cases, the orbits
have eccentricity $e = 0.7$, inclination $\iota = 60^\circ$, and are
about a hole with spin $a = 0.999M$.  Even deep in the strong field,
$L^2$ differs very little from $Q + L_z^2$ --- the greatest deviation
in this sample is about $1\%$.  Since $Q + L_z^2$ is a constant by
definition, this also shows that $L^2$ is approximately conserved over
the orbit.}
\label{fig5}
\end{figure}

\section{}
\label{app:formulae}

This appendix contains explicit expressions (in terms of $E$, $L_z$,
$Q$ and their derivatives) for the various functions appearing in the
formulae (\ref{rates}) for the rates $\dot p$, $\dot e$, $\dot\iota$.
First,
\begin{equation}
H = Q_{,p} E_{,e} L_{z,\iota} -Q_{,p} E_{,\iota} L_{z,e}
- Q_{,e}E_{,p}L_{z,\iota}
+ Q_{,e} E_{,\iota} L_{z,p} + Q_{,\iota} E_{,p} L_{z,e} 
- Q_{,\iota} E_{,e} L_{z,p} \;.
\label{Hfunc}
\end{equation}
For $\dot p$ we find
\begin{eqnarray}  
b_p &=& Q_{,\iota} L_{z,e} -Q_{,e} L_{z,\iota} \; ,
\\
c_p &=& E_{,\iota}Q_{,e} -E_{,e} Q_{,\iota} \;,
\\
d_p &=& E_{,e} L_{z,\iota} -E_{,\iota} L_{z,e} \;.
\end{eqnarray}
For $\dot{e}$,
\begin{eqnarray}
b_e &=& L_{z,\iota} Q_{,p} -Q_{,\iota} L_{z,p} \;,
\\
c_e &=&  Q_{,\iota} E_{,p} -E_{,\iota} Q_{,p} \;,
\\
d_e &=& E_{,\iota} L_{z,p} -E_{,p} L_{z,\iota} \;.
\end{eqnarray}
Finally, for $ \dot{\iota}$ the coefficients are,
\begin{eqnarray}
b_{\iota} &=& Q_{,e} L_{z,p} -Q_{,p} L_{z,e} \;,
\\
c_{\iota} &=& Q_{,p} E_{,e} -Q_{,e} E_{,p} \;,
\\
d_{\iota} &=& E_{,p} L_{z,e} -E_{,e} L_{z,p} \;.
\end{eqnarray}


\begin{minipage}[t]{6in}
\begin{table}
\begin{tabular}{ccccc}
$a/M$ & $e$ & $p_{\rm crit}/M $ (numerical) & $p_{\rm crit}/M$
(hybrid approx.) & Fractional difference \\
\hline
0     & 0.2  &  6.76   &  6.92  &  0.0237 \\
0     & 0.4  &  6.99   &  7.13  &  0.0200 \\
\hline
0.5   & 0.3  &  4.85   &  5.06  &  0.0433 \\
0.5   & 0.5  &  5.08   &  5.21  &  0.0250 \\
\hline
-0.99 & 0.3  & 10.25   & 10.53  &  0.0273 \\
-0.99 & 0.5  & 10.59   & 10.78  &  0.0179 \\
\end{tabular}
\caption[Table]
{Comparing critical curve values $p_{\rm crit}$ for equatorial
eccentric orbits.  These are the $p$ values at which the eccentricity
evolution switches sign, beginning to grow rather than shrink.  We
show the values of $p_{\rm crit}$ calculated numerically
{\cite{cutler,kgdk}} (third column) and using the hybrid approximation
(fourth column), for a variety of black hole spins (negative $a/M$
represents retrograde orbits) and eccentricities.  The fifth column
shows the fractional difference between the numerical and the
approximate results, (\mbox{approximate}
-\mbox{numerical})/(\mbox{numerical}).}
\label{tab1}
\end{table}
\end{minipage}

\begin{minipage}[t]{6in}
\begin{table}

\begin{tabular}{cccccccc}
  
$a/M$ & $p/M$ & $e$ & Calculation & $(M/\mu)\dot{p}$ & Frac.\ diff.\
in $\dot p$ & $(M^2/\mu)\dot{e}$ & Frac.\ diff.\ in $\dot e$ \\
\hline
0 & 7.505 & 0.189 & Numerical     & $-7.475\times10^{-2}$ &  ---   &
$-1.967\times10^{-3}$ &  ---   \\
 &  &  & Hybrid        & $-6.859\times10^{-2}$ & 0.0824 &
$-1.291\times10^{-3}$ & 0.3434 \\
 &  &  & Leading order & $-2.957\times10^{-2}$ & 0.6044 &
$-1.159\times10^{-3}$ & 0.4108 \\
\hline
0 & 6.9 & 0.4 & Numerical     & $-4.240\times 10^{-1} $& --- &
$ + 1.047 \times 10^{-2} $ & --- \\
 &  &  & Hybrid    &  $-3.056\times 10^{-1} $  & 0.2792    & 
$+1.506\times10^{-2}$  & -0.4384 \\
 &  &  & Leading order    & $ -3.420\times10^{-2}$   & 0.9193  &
$ -2.929\times10^{-3}$  & 1.2797 \\
\hline
0.5 & 6.5 & 0.4 & Numerical     & $-5.999\times10^{-2}$ &  ---   &
$-5.155\times10^{-3}$ &  ---   \\
 &  &  & Hybrid        & $-4.606\times10^{-2}$ & 0.2322 &
$-3.356\times10^{-3}$ & 0.3490 \\ 
 &  &  & Leading order & $-4.091\times10^{-2}$ & 0.3181 &
$-3.719\times10^{-3}$ & 0.2786 \\
\hline
0.5 & 15 & 0.4 & Numerical     & $-3.371\times10^{-3}$ &  ---   &
$-1.341\times10^{-4}$ &  ---   \\
 &  &  & Hybrid        & $-3.358\times10^{-3}$ & 0.0039 &
$-1.334\times10^{-4}$ & 0.0052 \\
 &  &  & Leading order & $-3.328\times10^{-3}$ & 0.0128 &
$-1.311\times10^{-4}$ & 0.0224 \\
\hline
0.5  & 4.8   & 0.3 & Numerical     & $-6.354\times 10^{-1}$ &  ---   &
$+1.369\times10^{-2}$ & --- \\
  &    &  & Hybrid     & $-4.858\times10^{-1} $ & 0.2354    &
$ +3.519\times10^{-2} $ & -1.5705    \\
  &    &  & Leading order & $-4.849\times10^{-2} $ & 0.9237 &
$-4.432\times10^{-3} $ & 1.3237    \\
\hline
0.9  & 5   & 0.4 & Numerical     & $-7.507\times 10^{-2}$ &  ---    &
$-9.266\times 10^{-3}$ & --- \\
  &    &  & Hybrid     & $ -4.617\times10^{-2} $ & 0.3850    &
$ -1.965\times10^{-3}$ & 0.7879  \\
  &    &  & Leading order & $ -2.698\times10^{-3} $ & 0.9641 &
$-9.732\times10^{-5}$  & 0.9895    \\
\hline
-0.5  & 10   & 0.4 & Numerical     & $-3.115\times10^{-2}$ &  ---   &
$-1.379\times 10^{-3}$ & --- \\
  &   &  & Hybrid     & $-2.494\times10^{-2} $ & 0.1993    &
$ -9.107\times10^{-4}$ & 0.3396    \\
  &    &  & Leading order & $-1.337\times10^{-2} $ & 0.5708 &
$-7.931\times10^{-4}$ & 0.4249    \\
\hline
-0.99 & 10.5 & 0.4 & Numerical     & $-7.516\times10^{-2}$ &  ---   &
$-5.312\times10^{-4}$ &  ---   \\
 &  &  & Hybrid        & $-5.506\times10^{-2}$ & 0.2674 &
$-5.223\times10^{-4}$ & 0.0168 \\
 &  &  & Leading order & $-9.704\times10^{-3}$ & 0.8709 &
$-5.461\times10^{-4}$ & 0.0281 \\
\hline
-0.99 & 15 & 0.4 & Numerical     & $-5.766\times10^{-3}$ &  ---   &
$-2.141\times10^{-4}$ &  ---   \\
 &  &  & Hybrid        & $-5.295\times10^{-3}$ & 0.0817 &
$-1.875\times10^{-4}$ & 0.1242 \\
 &  &  & Leading order & $-3.328\times10^{-3}$ & 0.4228 &
$-1.311\times10^{-4}$ & 0.3877 \\
\end{tabular}
\caption[Table]
{Comparing the rates $\dot{p}$, $\dot{e}$ for several equatorial
eccentric orbits.  The fifth column in this table shows $\dot p$; the
seventh column shows $\dot e$.  Within each section of the table, the
first row of columns five and seven contains accurate numerical data
from {\cite{cutler,kgdk}}, the second row shows data using the hybrid
scheme outlined in this paper, and the third row shows data using
quadrupole order results.  The sixth and eighth columns show the
fractional differences between the two approximation schemes and the
accurate numerical results.  In all cases but one, the hybrid
approximation is closer to the accurate numerical calculation,
sometimes substantially so.}
\label{tab2}
\end{table}
\end{minipage}


\begin{minipage}[t]{6in}
\begin{table}
\begin{tabular}{cccccccc}  
$a/M$ & $p/M$ & $\iota({\rm degrees})$ & Calculation &
$(M/\mu)\dot{p}$ & Frac.\ diff.\ in $\dot p$ & $(M^2/\mu) \dot{\iota}$
& Frac.\ diff.\ in $\dot\iota$ \\
\hline
0.95 & 7 & 62.43 & Numerical     & $-4.657\times10^{-2}$ &  ---   &
$1.207\times10^{-4}$ &  ---   \\
     &   &       & Hybrid        & $-4.497\times10^{-2}$ & 0.0344 &
$2.639\times10^{-4}$ & 1.1864 \\
     &   &       & Leading order & $-2.750\times10^{-2}$ & 0.4095 &
$3.080\times10^{-4}$ & 1.5518 \\
\hline
0.05 & 7 & 60.17 & Numerical     & $-1.096\times10^{-1}$ &  ---   &
$1.087\times10^{-5}$ &  ---   \\
     &   &       & Hybrid        & $-1.048\times10^{-1}$ & 0.0438 &
$1.207\times10^{-5}$ & 0.1104 \\
     &   &       & Leading order & $-3.676\times10^{-2}$ & 0.6642 &
$1.587\times10^{-5}$ & 0.4500 \\
\hline
0.5     & 10 & 67.56  & Numerical    & $-1.583\times10^{-2} $ & --- &
$ 1.546\times10^{-5} $ & --- \\
     &  &   &  Hybrid    & $ -1.645\times10^{-2}  $ & 0.0392 &
$ 2.043\times10^{-5} $ & 0.3215  \\
     &   &  & Leading order & $ -1.194\times10^{-2} $ & 0.2457  &
$ 2.377\times10^{-5} $ & 0.5375 \\
\hline
0.5     & 10 & 126.76   & Numerical    & $ -2.329\times10^{-2}  $ & --- &
$ 1.892\times10^{-5}  $ & --- \\
     & &  & Hybrid    & $ -2.341\times10^{-2} $ & 0.0051  &
$ 1.643\times10^{-5} $ & 0.1316  \\
     &  &  & Leading order    & $ -1.414\times10^{-2} $ & 0.3929  &
$ 2.060\times10^{-5} $ & 0.0888 \\
\hline
0.9     & 10 & 74.07 & Numerical    & $ -1.544\times10^{-2}  $ & --- &
$ 2.715\times10^{-5} $ & --- \\
     & &  & Hybrid    & $  -1.567\times10^{-2} $ & 0.0149 &
$ 3.857\times10^{-5} $ & 0.4206 \\
     & &  & Leading order    & $-1.169\times10^{-2} $ & 0.2429 &
$ 4.452\times10^{-5} $ & 0.6398 \\
\hline
0.9     & 10 & 131.57    & Numerical    & $ -3.253\times10^{-2}  $ & --- &
$ 3.887\times10^{-5}  $ & --- \\
     &  & & Hybrid    & $ -3.082\times10^{-2}  $ & 0.0526 &
$ 2.612\times10^{-5} $ & 0.3280 \\
     &  &  & Leading order    & $ -1.545\times10^{-2} $ & 0.5250 &
$ 3.464\times10^{-5} $ & 0.1088 \\
\hline
0.5     & 6  & 48.33     & Numerical    & $ -1.237\times10^{-1}   $ & --- &
$ 1.410\times10^{-4} $ & --- \\
     &  & & Hybrid    & $ -1.135\times10^{-1} $ & 0.0824 &
$ 2.614\times10^{-4} $ & 0.8539 \\
     &  &  & Leading order    & $ -4.440\times10^{-2} $ & 0.6411 &
$ 3.190\times10^{-4} $ & 1.2624 \\
\hline
0.5     & 6 &  67.81  & Numerical    & $ -2.020\times10^{-1} $ & --- &
$ 2.094\times10^{-4}$ & --- \\
     &  & & Hybrid    & $ -1.779\times10^{-1}   $ & 0.1193 &
$ 2.992\times10^{-4}  $ & 0.4288 \\
     &  &  & Leading order    & $ -5.082\times10^{-2} $ & 0.7484 &
$ 3.954\times10^{-4} $ & 0.8882 \\
\hline
0.9     & 6  & 54.64     & Numerical    & $ -7.846\times10^{-2}  $ & --- &
$ 2.015\times10^{-4}  $ & --- \\
     &  & & Hybrid    & $ -6.950\times10^{-2} $ & 0.1142 &
$ 5.486\times10^{-4} $ & 1.7226 \\
     &  &  & Leading order    & $ -3.598\times10^{-2} $ & 0.5674 &
$ 6.268\times10^{-4} $ & 2.1107 \\
\hline
0.9     & 6 & 99.55   & Numerical    & $ -74.32 $ & --- &
$ 6.337\times10^{-4}  $ & --- \\
     &  & & Hybrid    & $ -48.02 $ & 0.3539 &
$ 5.241\times10^{-4}  $ & 0.1729 \\
     &  &  & Leading order    & $ -6.593\times 10^{-2} $ & 0.9991 &
$ 7.580\times10^{-4} $ & 0.1961 \\
\end{tabular}
\caption[Table]
{Comparing the rates $\dot{p}$, $\dot{\iota}$ for several inclined
circular orbits.  The fifth column shows $\dot p$; the seventh column
shows $\dot\iota$.  Within each section of the table, the first row of
columns five and seven contains accurate numerical data from
{\cite{scott1}}, the second row shows data using the hybrid scheme
outlined in this paper, and the third row shows data using leading
order results.  The sixth and eighth columns show the fractional
differences between the two approximation schemes and the accurate
numerical results. In most cases in this sample, the hybrid scheme
performs much better than the leading-order approximation when
compared to the rigorous numerical data.  The only case which this is
not true is for $\dot{\iota}$ of retrograde orbits.  Nevertheless,
this small inaccuracy has no impact on the calculation of generic
inspirals, as we assume that $\iota=\mbox{constant}$.  Note the
enormous difference between the numerical and the leading order
results in the Table's final entry.  This is because that point is
fairly close to the separatrix between stable and unstable orbits.
Since the leading-order calculation has no notion of this separatrix,
it is particularly inaccurate here.}
\label{tab3}
\end{table}
\end{minipage}


\end{document}